\documentclass[12pt]{article}
\usepackage{amsmath, amsfonts, amsthm, amssymb, amscd}
\usepackage[mathcal]{euscript}
\usepackage{color}
\usepackage{fullpage}
\usepackage{ulem}

\newcounter{mnotecount}[section]

\begin{document}

\vspace{-2cm}

\title{Non-rigidly rotating stationary cylindrical dust spacetimes}

\author{R. Chan$^1$\thanks{e-mail: chan@on.br} and N. O. Santos$^2$\thanks{e-mail: nilton.santos@obspm.fr}\\
{\small $^1$ Coordena\c{c}\~ao de Astronomia e Astrof\'{\i}sica, Observat\'orio Nacional,}\\
{\small Rua General Jos\'e Cristino 77, S\~ao Crist\'ov\~ao, 20921-400, Rio de Janeiro, Brazil.}\\
{\small $^2$ Sorbonne Universit\'e, UPMC Universit\'e Paris 06, LERMA, Observatoire de Paris-Meudon,}\\
{\small 5 place Jules Janssen, F-92195 Meudon Cedex, France.}}
\maketitle

\begin{abstract}
We consider stationary rotating cylindrically symmetric dust spacetimes. We first show
that the Maitra spacetime is the unique non-rigidly 
(non null shear scalar) rotating solution with a regular axis 
and that is the most general one of the field
equations. We are also able to demonstrate what Maitra's paper does not show, where 
the solution is merely stated without any demonstration.
Then we find that the non-rigidly rotating spacetimes, not necessarily regular, can be matched, across timelike cylindrical hypersurfaces, to a one-parameter family of stationary vacuum exteriors given by the Weyl class but not to the Lewis class, generalizing a result by Bonnor and Steidman. 
Among other properties, it is shown that the amount of rotating dust packed inside a cylinder is bigger if it is rigidly rotating than non-rigidly rotating. 
Finally, we show that in the non-rigidly rotating cases the fluid necessarily has vorticity and the spacetimes are not silent, where the magnetic part of the Weyl tensor is non-zero.

\end{abstract}
\newpage

\section{Introduction}

Exact solutions to the Einstein equations provide idealized yet valuable models to study physical phenomena in General Relativity. In this context, cylindrically symmetric exact solutions have been used to study e.g. gravitational collapse \cite{Thorne, Brito} and gravitational radiation \cite{Einstein,Ehlers,Stephani,Griffiths2009,Astekar,Garcia-Mena}. The importance of cylindrical systems in General Relativity has been revised in \cite{Bronnikov}.

Among the cylindrical solutions, the ones which are stationary rotating are of particular interest for local models in astrophysics. The stationary cylindrically symmetric vacuum solutions are the well-known Lewis spacetime \cite{Exact-Book}. Some rigidly rotating (i.e. null shear scalar) perfect fluid particular solutions are also known for a long time (see Section 22.2 of \cite{Exact-Book}) but, interestingly, an exhaustive investigation has recently been performed by Celerier who found new rigidly rotating solutions and explored some of their properties \cite{Krasinski1975,Krasinski1978,Celerier1, Celerier2}. 

Exact solutions of non-rigidly rotating cylindrical fluid models can also be found in \cite{Exact-Book}, see Section 22.2, including the Maitra solution containing dust matter \cite{Maitra}. We call attention that this solution was not obtained in the 
Maitra's original paper but merely cited without any demonstration.
This solution is specially interesting as it is non-singular at the axis, is asymptotically flat, does not contain timelike curves and is geodesically complete. Furthermore the Maitra solution has been matched to a vacuum exterior by Bonnor and Steadman \cite{Steadman}, who showed that not all the Lewis  spacetime can act as exteriors to Maitra's solution restricting this possibility to the Weyl class of solutions only.  
Non-rigidly rotating models have also been considered by \cite{Bratek} and very recently, more systematically, by Celerier in \cite{Celerier6}, who considered irrotational matter only.

The general stationary axially symmetric for non rigidly rotating dust was founded, in 1975, by Winicour \cite{Winicour}. The solution was not presented explicitly, depending upon one arbitrary function of one variable and in addition to an arbitrary solution of a Laplace equation. The case of rigid rotation was solved long before, in 1937, by van Stockum \cite{Stockum,Lanczos}, with a solution being depended of one arbitrary function of the Laplace equation.
The specialization to the cylindrical case has been given by Vishveshwara and Winicour \cite{Vishveshwara} in 1977, which it is not the aim of the present work.
In the van Stockum solution dust is rigidly rotating. Of course, it is interesting to see what would imply non rigid on the rotating dust. Such a solution was cited but not obtained by Maitra \cite{Maitra} in 1966, which has a special interest since it has a regular axis of symmetry. Later King \cite{King}, in 1974, re-obtained the cited Maitra's solution.

In this paper, we take this effort further and consider non-rigidly rotating cylindrically symmetric models having matter with vorticity. For simplicity, and a starting point, we look for solutions with dust matter. After imposing regularity conditions at the axis, 
we recover the Maitra solution which we show that it is the most general 
one of the field
equations. We are also able to demonstrate what Maitra's paper does not show, where 
the solution is merely stated without any demonstration.
We then study the junction of general non-rigidly rotating dust spacetimes to stationary vacuum exteriors and investigate some physical properties comparing to the rigidly rotating case.
We find that cylindrically symmetric non-rigidly rotating dust matter cannot be matched to an arbitrary vacuum cylindrical stationary 
spacetime belonging to the Lewis class but only to the Weyl class, thus generalizing the result of \cite{Steadman}. This fact is interesting since only the Weyl class has a static Levi Civita vacuum Newtonian limit.

Dust matter is abundant in the universe and there are still unsettled questions about the quantification of the gravitational energy associated to this type of matter in which the Weyl tensor plays an important role  (e.g. for locally cylindrical systems see \cite{Garcia-Mena,MacCallum-Santos,Tod-Mena}). 
It is then interesting to study the Weyl tensor for stationary rotating dust spacetimes.  Dust matter is {\em silent} in the typical cosmological models. {\em Silent spacetimes} are dust spacetimes such that the fluid velocity vector is irrotational and the magnetic part of the Weyl tensor vanishes. This definition was put forward by \cite{Matarrese} in order to study models of structure formation and there is a considerable amount of work, as well as conjectures, about the existence of silent spacetimes (see e.g. \cite{Mars,Norbert} and references therein). Rotation can play a role in this debate since it can trigger a non-zero magnetic part of the Weyl tensor which, in turn, is a purely relativistic effect as compared to Newtonian dust. This will also be explored in this sequel. 

The paper is organized as follows: Section \ref{intro} contains a short introduction to stationary cylindrically symmetric spacetimes, while Section \ref{dust} particularizes to spacetimes containing dust matter and then to solutions regular at the axis of symmetry. In turn, in Section \ref{f=1} we explore explicit  dust solutions in the proper time gauge. Section \ref{matching} is dedicated to the study of matching interior non-rigidly rotating stationary dust sources to vacuum exteriors and Section \ref{Weyl} to the Weyl tensor and its relation to rotation. 

\section{Cylindrically symmetric stationary spacetimes}
\label{intro}

In this section we revise some basic properties of cylindrical stationary spacetimes, not necessarily dust, and the expressions of useful kinematic quantities. More details are contained in \cite{Celerier}.

We assume a general stationary cylindrically symmetric spacetime metric
\begin{equation}
ds^2=-fdt^2+2kdtd\phi+e^{\gamma}(dr^2+dz^2)+ld\phi^2, \label{3}
\end{equation}
where $f$, $k$, $\gamma$ and $l$ are all functions of $r$ and  the following ranges on the coordinates
\begin{equation}
-\infty< t <\infty, \;\;  r\ge 0, \;\; -\infty<z<\infty, \;\; 0\leq\phi < 2\pi. \label{4}
\end{equation} 
We number the coordinates as $x^0=t$, $x^1=r$, $x^2=z$ and $x^3=\phi$. 
\footnote{The metric equation (22.4a) \cite{Exact-Book} is used because
we assume the metric functions are real and the coordinate $\phi$ is a angular coordinate.}
We consider fluid matter distribution with a timelike four-velocity $V_{\alpha}$ satisfying
\begin{equation}
 V_{\alpha}V^{\alpha}=-1. \label{2}
\end{equation}
The fluid is assumed to rotate non-rigidly, in general, and its four velocity  can be expressed  as
\begin{equation}
V^{\alpha}=v\delta^{\alpha}_0+\Omega\delta^{\alpha}_3, \label{4a}
\end{equation}
where $v$ and $\Omega$ are functions of $r$ only, and since it satisfies the timelike condition (\ref{2}) we have
\begin{equation}
fv^2-2kv\Omega-l\Omega^2-1=0, \label{5a}
\end{equation}
or
\begin{equation}
fv=\epsilon(f+D^2\Omega^2)^{1/2}+k\Omega, \label{6a}
\end{equation}
where $\epsilon=\pm 1$ and
\begin{equation}
D^2=fl+k^2. \label{7a}
\end{equation}
One can split the spacetime in a $1+3$ slicing relative to $V^\alpha$ where the expansion associated to $V^\alpha$ is zero due to stationarity. In that case, we can decompose the covariant derivative of $V^\alpha$ as:
\begin{equation}
\nabla _{\beta}V_{\alpha}=-{\dot V}_{\alpha}V_{\beta}+\omega_{\alpha\beta}+\sigma_{\alpha\beta}, \label{8a}
\end{equation}
where the kinematic quantities defined as
\begin{eqnarray}
\aligned
{\dot V}_{\alpha}&=V^{\beta}\nabla_{\beta}V_{\alpha},
\\
\omega_{\alpha\beta} &=\nabla_{[\beta}V_{\alpha]}+{\dot V}_{[\alpha}V_{\beta]}, 
\\
\sigma_{\alpha\beta}&=\nabla_{(\beta}V_{\alpha)}+{\dot V}_{(\alpha}V_{\beta)} \label{11aa}
\endaligned
\end{eqnarray}
are the acceleration, vorticity and shear, respectively. With (\ref{4a}) the non-zero components of (\ref{11aa}) are
\begin{eqnarray}
\aligned
{\dot V}_1&=-\Psi, \label{12a}\\
2\omega_{01}&=-(fv-k\Omega)^{\prime}-(fv-k\Omega)\Psi, \label{13a}\\
2\omega_{13}&=-(kv+l\Omega)^{\prime}-(kv+l\Omega)\Psi, \label{14a}\\
2\sigma_{01}&=D^2\Omega(\Omega v^{\prime}-v\Omega^{\prime}), \label{15a}\\
2\sigma_{13}&=-D^2v(\Omega v^{\prime}-v\Omega^{\prime}), \label{16a}
\endaligned
\end{eqnarray}
where the prime stand for differentiation with respect to $r$ and
\begin{equation}
2\Psi=-v^2f^{\prime}+2v\Omega k^{\prime}+\Omega^2l^{\prime}, \label{17a}
\end{equation}
or by using (\ref{5a}) 
\begin{equation}
\Psi=v(fv^{\prime}-k\Omega^{\prime})-\Omega(kv^{\prime}+l\Omega^{\prime}). \label{17aa}
\end{equation}
In turn, the norm of each quantity in (\ref{11aa}) becomes
\begin{eqnarray}
{\dot V}^{\alpha}{\dot V}_{\alpha}&=&e^{-\gamma}\Psi^2, \\
\omega_{\alpha\beta}\omega^{\alpha\beta}&=&\frac{1}{2D^2e^{\gamma}v^2}[(kv+l\Omega)^{\prime}+(kv+l\Omega)\Psi]^2, \label{19a}\\
\sigma_{\alpha\beta}\sigma^{\alpha\beta}&=&\frac{D^2(\Omega^{\prime}-\Omega\Psi)^2}{2e^{\gamma}(f+D^2\Omega^2)}. 
\end{eqnarray}
\section{Cylindrically symmetric rotating dust matter}
\label{dust}

In this section we write the Einstein equations for the case of dust matter, which form a particular case of the equations in \cite{Celerier} for more general fluids and explore some consequences. 
%
\subsection{Field equations and general properties}

We consider an energy momentum tensor given by
\begin{equation}
T_{\alpha\beta}=\mu V_{\alpha}V_{\beta}, \label{x1}
\end{equation}
where $\mu=\mu(r)$ is the matter density of the rotating dust fluid with timelike four velocity $V^{\alpha}$ satisfying (\ref{5a}).
The Einstein field equations, $G_{\alpha\beta}=\kappa T_{\alpha\beta}$, with (\ref{3}), (\ref{4a}) and (\ref{x1}) give the non-zero components:
\begin{align} 
G_{00}&=-\frac{1}{4D^2e^{\gamma}}\left[2fD^2\gamma^{\prime\prime}+4fDD^{\prime\prime}-2D^2f^{\prime\prime}+2DD^{\prime}f^{\prime} \right. 
\left.-3f(f^{\prime}l^{\prime}+k^{\prime 2})\right]= \kappa\mu(f+D^2\Omega^2), \label{5} \\
G_{03}&=\frac{1}{4D^2e^{\gamma}}\left[2kD^2\gamma^{\prime\prime}+4kDD^{\prime\prime}-2D^2k^{\prime\prime}+2DD^{\prime}k^{\prime} \right.
\left. -3k(f^{\prime}l^{\prime}+k^{\prime 2})\right]=-\kappa\mu(k+D^2v\Omega), \label{6} \\
G_{11}&=\frac{1}{4D^2}(2DD^{\prime}\gamma^{\prime}+f^{\prime}l^{\prime}+k^{\prime 2})=0, \label{7} \\
G_{22}&=-\frac{1}{4D^2}(2DD^{\prime}\gamma^{\prime}-4DD^{\prime\prime}+f^{\prime}l^{\prime}+k^{\prime 2})=0, \label{8} \\
G_{33}&=\frac{1}{4D^2e^{\gamma}}\left[2lD^2\gamma^{\prime\prime}+4lDD^{\prime\prime}-2D^2l^{\prime\prime}+2DD^{\prime}l^{\prime} \right. 
\left. -3l(f^{\prime}l^{\prime}+k^{\prime 2})\right]=-
\kappa\mu(l-D^2v^2). \label{9}
\end{align}
Adding (\ref{7}) and (\ref{8}) we obtain $D^{\prime\prime}=0$, which after integration becomes
\begin{equation}
D=c_1r+c_2, \label{10}
\end{equation}
where $c_1$ and $c_2$ are integration constants.
From (\ref{7}) or  (\ref{8}) we obtain
\begin{equation}
\gamma^{\prime}=-\frac{1}{2DD'}(f^{\prime}l^{\prime}+k^{\prime 2}). \label{13b}
\end{equation}
In turn, the Bianchi identities $\nabla_\beta T^{~\beta}_{\alpha}=0$, with (\ref{3}), (\ref{4a}) and (\ref{x1}) reduce to
$
\nabla_\beta T^{~\beta}_{1}=-\mu\Psi=0, \label{20aa}
$
which implies
\begin{equation}
\Psi=0, \label{21aa}
\end{equation}
so that acceleration is zero as expected for dust matter, and
\begin{align}
\omega_{\alpha\beta}\omega^{\alpha\beta}&=\frac{1}{2e^{\gamma}D^2} \left[\epsilon(-2DD'\gamma^{\prime})^{1/2}+
\frac{D^2\Omega^{\prime}}{(f+D^2\Omega^2)^{1/2}}\right]^2, \label{ww}\\
\sigma_{\alpha\beta}\sigma^{\alpha\beta}&=\frac{D^2\Omega^{\prime 2}}{2e^{\gamma}(f+D^2\Omega^2)}.
\end{align}
The Raychaudhuri equation for stationary cylindrical dust gives just
\begin{equation}
\frac{\kappa}{2}\mu=\omega_{\alpha\beta}\omega^{\alpha\beta}-\sigma_{\alpha\beta}\sigma^{\alpha\beta}, \label{23aa}
\end{equation}
which suggests that the presence of shear, producing non-rigid rotation, reduces the amount of matter density that can be supported by a rotating dust cylinder.

A general useful expression to calculate $\Omega$ in terms of the metric (\ref{3}) and its derivatives can be obtained from (\ref{6a}) and (\ref{21aa}) as
\begin{equation}
\label{exp1}
2D(Df^{\prime}-fD^{\prime})\Omega^2+2\epsilon (kf^{\prime}-fk^{\prime})(f+D^2\Omega^2)^{1/2}\Omega+ff^{\prime}=0. 
\end{equation}
\subsection{Regularity and Whittaker mass}
We now impose regularity conditions at the axis of symmetry. Geometrically the regularity condition at the axis can be translated into the flatness condition \cite{Exact-Book}
\begin{equation}
\lim_{r\to 0}\frac{\nabla_\beta(X^\alpha X_\alpha) \nabla^\beta(X^\mu X_\mu)}{4X^\nu X_\nu}=1
\end{equation}
where $X^\alpha=\delta^\alpha_{\phi}$ is the infinitesimal generator of the axial symmetry and is spacelike in a neighborhood of the axis. This condition ensures the $2\pi$-periodicity of the axial coordinate around the axis. Physically along the axis there cannot exist singularities, hence in our case the regularity conditions on the metric coefficients give
\begin{equation}
f=e^{\gamma}=1, \;\; k=l=0, \;\; f^{\prime}=\gamma^{\prime}=k^{\prime}=l^{\prime}=0, ~~~{\text {at~$r=0.$}} \label{11}
\end{equation}
Hence, we have from (\ref{10}), assuming $c_1=1$ and $c_2=0$,
\begin{equation}
D=r, \label{12}
\end{equation}
where we have rescaled $r$.  The condition $c_2=0$ comes from the regularity
condition and in the condition $c_1=1$ we use the gauge freedom of redefinition of the radial 
coordinate. 

We now calculate the Whittaker active mass per unit length \cite{Whittaker} of a cylindrical stationary distribution of dust with energy density $\mu(r)$ which is given by 
\begin{equation}
M(r)=2\pi\int^r_0\mu\sqrt{-g}dr. \label{1x}
\end{equation}
We can obtain $\mu(r)$ from  (\ref{x1}), (\ref{5}), (\ref{6}) and (\ref{9}) as
\begin{equation}
g^{\alpha\beta}T_{\alpha\beta}=-\kappa\mu=\frac{(r\gamma^{\prime})^{\prime}}{re^{\gamma}}, \label{2x}
\end{equation}
and with (\ref{11}) the integration (\ref{1x}) results
\begin{equation}
M(r)=-\frac{r}{4}\gamma^{\prime}. \label{3x}
\end{equation}
Now with (\ref{13b}), \eqref{exp1} and (\ref{12}) we get
\begin{equation}
(2r\gamma^{\prime}+1)r^2\Omega^4+(2f\gamma^{\prime}+f^{\prime})r\Omega^2+\frac{f^{\prime 2}}{4}=0. \label{41bb}
\end{equation}
Substituting (\ref{3x}) into (\ref{41bb}) we obtain for $\Omega\neq 0$,
\begin{equation}
M(r)=\frac{(2r^2\Omega^2+rf^{\prime})^2}{32r^2\Omega^2(r^2\Omega^2+f)} \label{42bb}
\end{equation}
from which we see how the mass density $M$ depends on the kinetic rotating energy of the dust through its tangential rotating velocity $r\Omega$. 
\section{Family of regular solutions in proper time}
\label{f=1}

In the metric \eqref{3} we can always choose the time coordinate to coincide with proper time
$\tau$ (making the transformation $d \tau=f dt$) in which case 
\begin{equation}
\label{f-equ}
f=1.
\end{equation}
Then equation (\ref{41bb}) gives the following two classes of solutions. 

\subsection{Rigidly rotating case $\Omega=0$}

For this class we have that the shear vanishes, $\sigma_{\alpha\beta}=0$, and the system is in a rigidly co-rotating frame.
The solution of the field equations (\ref{5})-(\ref{9}) for this class and satisfying the regularity conditions (\ref{11}) is 
\begin{equation}
f=1, \;\; k=\omega r^2, \;\; l=r^2(1-\omega^2r^2), \;\; \gamma=-\omega^2 r^2, ~~v=1\label{z3}
\end{equation}
and for the matter density
\begin{equation}
\kappa\mu=4\omega^2e^{\omega^2r^2}, \label{z4}
\end{equation}
where $\omega$ is a constant.
This solution was obtained by van Stockum \cite{Stockum} and studied by many authors (for a review  see \cite{Bronnikov}).
Its Whittaker active mass becomes
\begin{equation}
M(r)=\frac{1}{2}\omega^2 r^2, \label{j13}
\end{equation}
which shows that it is associated to the Newtonian kinetic energy of the rotating dust with tangential velocity $\omega r$. 

\subsection{Non-rigidly rotating case $\Omega=\Omega(r)$}
For this class, from equation (\ref{41bb}) we have
\begin{equation}
\label{Omega-exp}
\Omega^2=-\frac{2\gamma^{\prime}}{r(1+2r\gamma^{\prime})}
\end{equation}
where 
$\gamma^{\prime}\le 0$. In this case the shear does not vanish and rotation is non-rigid. 
In order to obtain the explicit metric functions for this class we proceed as follows. 

Note first that the cases $l-D^2 v^2=0$ (i.e. $G_{03}=0$) or $k+D^2v\Omega=0$ (i.e. $G_{33}=0$) do not give viable dust solutions, so we exclude them. 

Then by calculating $G_{00}/G_{03}$ and $G_{00}/G_{33}$ and using (\ref{5}), (\ref{6}) and (\ref{9}) we obtain respectively
\begin{eqnarray}
\frac{\Delta}{k\Delta-rk^{\prime\prime}+k^{\prime}}=\frac{1+r^2\Omega^2}{k+r^2v\Omega}, \label{j1} \\
\frac{\Delta}{l\Delta-rl^{\prime\prime}+l^{\prime}}=\frac{1+r^2\Omega^2}{l-r^2v^2}, \label{j2}
\end{eqnarray}
where
\begin{equation}
\Delta=r\gamma^{\prime\prime}+3\gamma^{\prime}. \label{j3}    
\end{equation}
For $l$ we obtain from \eqref{7a} and (\ref{12})
\begin{equation}
l=r^2-k^2, \label{j8}
\end{equation}
and we calculate $k$ next.

Considering (\ref{13b}) and transforming away $\Omega$ from (\ref{j1}) and (\ref{j2}) we obtain
\begin{equation}
rk^{\prime\prime}=\frac{(1-k^{\prime 2})k^\prime}{1+k^{\prime 2}}, \label{j4}    
\end{equation}
whose solution is
\begin{equation}
k^{\prime}=\frac{\varepsilon+(1+x^2)^{1/2}}{x}, \label{j5}    
\end{equation}
where $\varepsilon=\pm 1$, $c_1$ is an integration constant and $x=2c_1r$. Integrating (\ref{j5}) we obtain
\begin{eqnarray}
k=\frac{1}{2c_1}\left\{(1+\varepsilon)\ln x-\ln[1+(1+x^2)^{1/2}]\right. \nonumber 
\left.+(1+x^2)^{1/2}-\ln 2+c_2\right\}, \label{j6}
\end{eqnarray}
where $c_2$ is an integration constant. In order to see if $k$ and $k^{\prime}$ satisfy the regularity conditions (\ref{11}) we take the limit $r\rightarrow 0$ and see that we need $c_2=-1+2\ln 2$ and $\varepsilon =-1$. To check if (\ref{j5}) satisfy the regularity conditions one simply uses the l'Hospital rule. Hence the metric solution for $k$ is
\begin{equation}
k=\frac{1}{2c_1}\left\{(1+x^2)^{1/2}-\ln[1+(1+x^2)^{1/2}]-1+\ln 2\right\}. \label{j7}    
\end{equation}
For $\gamma$ we obtain from (\ref{13b}),
\begin{equation}
\gamma^{\prime}=-\frac{k^{\prime 2}}{2r}.   \label{j9} 
\end{equation}
By substituting (\ref{j5}) and integrating we find that the solution satisfying the regularity conditions (\ref{11}) reads
\begin{equation}
\gamma=\frac{1}{2}\left\{[1-(1+x^2)^{1/2}]\frac{1}{x^2}-\ln [1+(1+x^2)^{1/2}]+\ln 2+\frac{1}{2}\right\}. \label{j11}
\end{equation}
It turns out that metric \eqref{3} with $\gamma, k$ and $l$ given by \eqref{j11}, \eqref{j7} and \eqref{j8}, corresponds to the class of solutions found by Maitra \cite{Maitra}. 
We have then the following result:{\it 
The Maitra solution, re-obtained is this work as the most general solution, is the unique stationary non-rigidly rotating dust solution \eqref{3} satisfying the regularity conditions at the axis.} 
Notice that this result was obtained by direct integration of the field equations and this procedure clarifies the {\it ad hoc} assumptions made in \cite{Maitra}.

In this case the Whittaker mass (\ref{3x}) with (\ref{j11}) becomes
\begin{equation}
M(r)=\frac{1}{8x^2}[1-(1+x^2)^{1/2}]^2 \label{j12}
\end{equation}
and near the axis of symmetry, $r\ll 1$, we have
\begin{equation}
M(r) \cong \frac{1}{32} x^2,  \label{j14}
\end{equation}
which shows that $M=0$ along the axis of symmetry, satisfying a proper physical condition. Furthermore, (\ref{j14}) is equal to (\ref{j13}) if $c_1=4\omega$, showing that  in the neighborhood of the axis the anisotropic rotating dust is rigidly rotating. Asymptotically, for large values of $r$, we have from (\ref{j12}) $M\rightarrow 1/8$, which implies that: 
{\it Less mass can be packed in a non-rigidly rotating dust distribution than in its corresponding rigidly rotating case.}

 \section{Exterior stationary vacuum spacetime and junction conditions}
 \label{matching}

Consider now that the dust cylinders given by our original metric  \eqref{3}, which we now denote by $g^-$, have a finite radius. We wish to match this spacetime to a cylindrically symmetric stationary vacuum exterior given by the Lewis metric.
In fact this metric is usually presented with four parameters 
\cite{Bronnikov,Exact-Book,Griffiths} without being given a specific physical interpretation unless matched to a particular source.
%
\subsection{Stationary vacuum exterior}

We recall that the Lewis metric is given by  \cite{Lewis}
\begin{equation}
g^+=-Fdt^2+2Kdtd\phi+R^{\frac{n^2-1}{2}}(dR^2+dz^2)+Ld\phi^2, \label{25b}
\end{equation}
where
\begin{equation}
\aligned
F&=aR^{1-n}-a\delta^2R^{1+n},\\
K&=-(1-ab\delta)\delta R^{1+n}-abR^{1-n},\\
L&=\frac{(1-ab\delta)^2}{a}R^{1+n}-ab^2R^{1-n},
\endaligned
\end{equation}
with
\begin{equation}
\delta=\frac{c}{an}, \label{30b}
\end{equation}
and $a>0$, $b$, $c$ and $n$ are constant parameters, though not independent. The number of independent parameters is discussed in \cite{Malcolm} and their physical meaning in \cite{Costa}. The Lewis  spacetime contains two distinct classes of spacetimes depending on if $n$ is real, called Weyl class, or imaginary, called Lewis class. 

When $n$ is real, the remaining parameters are real too and the
metric can be put into a form where the Killing vector field $\partial_t$ is timelike,
and which depends only on three parameters with a clear physical significance: the Komar
mass, the angular momentum per unit length and the angle deficit \cite{Costa}.
The constant $n$ is related to the Newtonian energy per unit length $\sigma$ of the cylindrical source given by,
\begin{equation}
\sigma=\frac{1}{4}(1-n). \label{41d}
\end{equation}
This relation can be assumed since the Weyl class has the Newtonian limit.

When $n$ is imaginary, the remaining parameters have to be chosen such that they produce a real metric.
Here we do not elaborate more on the complex parameters as we shall not need them. For more information about the Lewis metric see e.g. \cite{Exact-Book, Costa}. Hereinafter, we will assume that $n$ is always real.

\subsection{Matching conditions}

We recall the matching conditions between two spacetimes $({M}^{\pm}, g^{\pm})$ across 
timelike hypersurfaces $\Sigma^{\pm}$ (see more details in \cite{Darmois, Mars-Seno}). The matching between 
two spacetimes requires an identification of their boundaries, i.e. a pair of embeddings
$\Phi^\pm:\; \Sigma \longrightarrow M^\pm$ with $\Phi^\pm(\Sigma) = \Sigma^{\pm}$, where $\Sigma$ is an abstract copy of one of the boundaries. Let $\xi^a$ be 
a coordinate system on $\Sigma$, where $a,b=1,2,3$. 
Given a vector basis $\{\partial/\partial \xi^a\}$ 
the push-forwards $d\Phi^\pm$ provide a correspondence between the vectors of this basis and sets of 
linearly independent vectors tangent to
$\Sigma^{\pm}$, given in appropriate coordinates by $e^{\pm \alpha}_a = \partial_{\xi^a} \Phi^{\pm \alpha}$, where we use the convention of previous sections $\alpha,\beta=0,1,2,3$. 
Consider also vectors $n_{\pm}^{\alpha}$ normal to the boundaries. 
The first and second fundamental forms on $\Sigma$ are given by 
 $q^\pm=\Phi^{\pm\star}(g^\pm)$ and $K^\pm=\Phi^{\pm\star}(\nabla^\pm {\bf n^\pm})$, where 
$\Phi^{\pm\star}$ denotes the pull-back corresponding to the maps $\Phi^\pm$. In components, those quantities can be written as 
\begin{equation}
q_{ab}^{\pm}\equiv e^{\pm \alpha}_a e^{\pm \beta}_b
g{}_{\alpha\beta}|_{{}_{\Sigma^\pm}},~~~
K_{ab}^{\pm}=-n^{\pm}_{\alpha} e^{\pm \beta}_a \nabla^\pm_\beta e^{\pm
\alpha}_b.
\end{equation}
The matching between two spacetimes through $\Sigma$ requires the equality of the first and second fundamental form on $\Sigma$, i.e.
\begin{equation}
  q_{ab}^{+}=q_{ab}^{-},~~~
  K_{ab}^{+}=K_{ab}^{-}.
\label{eq:backmc}
\end{equation}
In our case below, the spacetimes are matched across a cylinder of symmetry, so we choose the aforementioned basis using the corresponding Killing vectors restricted to $\Sigma$.

In turn the coordinates $t$, $z$ and $\phi$ can be taken as in (\ref{3}) and with the same ranges (\ref{4}). The radial coordinates in (\ref{3}) and (\ref{25b}), $r$ and $R$, are not necessarily continuous on $\Sigma$ in order
to preserve the range of $\phi$ in (\ref{4}). Also, since we are considering a cylinder without expansion, the boundary surface reads $r\stackrel{\Sigma}{=}$ constant and $R\stackrel{\Sigma}{=}$ constant, from inside and outside, respectively.

In accordance with the first junction conditions we obtain that the metric coefficients (\ref{3}) and (\ref{25b}) must be continuous across $\Sigma$ (see also \cite{Debbasch,Celerier})
\begin{equation}
f\stackrel{\Sigma}{=}F, \quad k\stackrel{\Sigma}{=}K, \quad \gamma \stackrel{\Sigma}{=}\Gamma, \quad  l\stackrel{\Sigma}{=}L, \label{31a}
\end{equation}
and the second matching conditions give
\begin{align}
\frac{f^{\prime}}{f}&\stackrel{\Sigma}{=}\frac{1}{R}+n\frac{\delta^2R^n+R^{-n}}{\delta^2R^{1+n}-R^{1-n}}, \label{32a}\\
\frac{k^{\prime}}{k}&\stackrel{\Sigma}{=}\frac{1}{R}+n\frac{(1-ab\delta)\delta R^n-abR^{-n}}{(1-ab\delta)\delta R^{1+n}+abR^{1-n}}, \label{33b}\\
\frac{l^{\prime}}{l}&\stackrel{\Sigma}{=}\frac{1}{R}+n\frac{(1-ab\delta)^2R^n+a^2b^2R^{-n}}{(1-ab\delta)^2R^{1+n}-a^2b^2R^{1-n}}. \label{35b}\\
\gamma^{\prime}&\stackrel{\Sigma}{=}\frac{n^2-1}{2R}, \label{34b}
\end{align}
The active mass per unit length $M(r)$ on the surface $\Sigma$ can be obtained from (\ref{3x}) and (\ref{34b}) as
\begin{equation}
M\stackrel{\Sigma}{=}\frac{1-n^2}{8}, \label{x4}
\end{equation}
or with the Newtonian mass per unit length $\sigma$ (\ref{41d}),
\begin{equation}
    M\stackrel{\Sigma}{=}\sigma(1-2\sigma), \label{x5}
\end{equation}
which does not depend on $\Omega$. 
But, in both cases, rigid and non-rigid rotation there is no static source limit, since by vanishing the rotation the spacetime becomes Minkowski. However, if the stationary source produces pressures, then the rotation  will contribute to the active mass.

We now focus on getting $\Omega$. From (\ref{5a}) and (\ref{21aa}) we obtain on $\Sigma$ respectively, by using (\ref{31a})-(\ref{35b}),
\begin{equation}
a(v+b\Omega)^2R^{1-n}-\frac{1}{a}\left[a\delta(v+b\Omega)-\Omega\right]^2R^{1+n}\stackrel{\Sigma}{=}1, \label{35c}
\end{equation}
and
\begin{equation}
(1-n)a^2(v+b\Omega)^2R^{-n}-(1+n)\left[a\delta(v+b\Omega)-\Omega\right]^2R^n\stackrel{\Sigma}{=}0, \label{35d}
\end{equation}
and by eliminating $v+b\Omega$ from both equations we obtain
\begin{equation}
\Omega^2\stackrel{\Sigma}{=}\frac{a}{2n}\left[(1-n)^{1/2}R^{-(1+n)/2}+\delta(1+n)^{1/2}R^{-(1-n)/2}\right]^2, \label{40b}
\end{equation}
The expression (\ref{40b}) shows us the interesting result that one needs 
$$\frac{a}{n}>0,$$
which means that $n$ has to be real and positive since $a>0$. Hence, while the range of $n$ for a rigidly rotating cylindrical dust matched to Weyl class spacetime is $1>n>-1$, the non-rigid rotating case has the range $1>n>0$. One can also say that for cylinders with finite radius, the dust matter under non-rigid rotation can pack less mass per unit length than a dust under rigid rotation, since for non-rigid dust rotation we have $0<\sigma<1/4$, while for rigid rotation $0<\sigma<1/2$.
Furthermore, the non-rigid rotation cannot be matched to the Lewis class, but only to the Weyl class of Lewis spacetimes. This agrees with the result found in \cite{Steadman} obtained for Maitra interiors and provides a generalization.

 \section{Weyl tensor and rotation}
\label{Weyl}

 In this section we will explore some properties of the Weyl tensor and rotation in the context of the previous sections.  The non-zero components of Weyl tensor for the metric \eqref{3} are presented in \cite{Celerier} and we rewrite them in Appendix.

\subsection{Electric and magnetic parts of the Weyl tensor}

The electric and magnetic parts of the Weyl tensor \cite{Exact-Book}, $E_{\alpha\beta}$ and $H_{\alpha\beta}$, respectively, are formed from the Weyl tensor $C_{\alpha\beta\gamma\delta}$ and its dual
$\tilde{C}_{\alpha\beta\gamma\delta}$ by contraction with the four velocity $V^{\alpha}$ as
\begin{equation}
\aligned
E_{\alpha\beta}&=C_{\alpha\gamma\beta\delta}V^{\gamma}V^{\delta}, \label{41b}\\
H_{\alpha\beta}&=\tilde{C}_{\alpha\gamma\beta\delta}V^{\gamma}V^{\delta}=\frac{1}{2}\epsilon_{\alpha\gamma\epsilon\delta}{C^{\epsilon\delta}}_{\beta\rho}V^{\gamma}V^{\rho},
\endaligned
\end{equation}
where $ \epsilon_{\alpha\beta\gamma\delta}\equiv\sqrt{-g}\,\eta_{\alpha\beta\gamma\delta}$ and $\eta_{\alpha\beta\gamma\delta}=+1$ or $-1$ for $\alpha,\beta,\gamma,\delta$ in even or odd order, respectively, and $\eta_{\alpha\beta\gamma\delta}=0$ otherwise.

Then it follows from equations (\ref{41b}) that the non-vanishing components of the electric part are
\begin{equation}
\aligned
E_{00}&=C_{0303}\Omega^2, \\
E_{03}&=-C_{0303}v\Omega,\\
E_{11}&=C_{0101}v^2-C_{0113}v\Omega, \label{45b}\\
E_{22}&=C_{0202}v^2-2C_{0223}v\Omega+C_{2323}\Omega^2, \\
E_{33}&=C_{0303}v^2,
\endaligned
\end{equation}
while the only non-zero component of the magnetic part is
\begin{eqnarray}
H_{12}=\frac{1}{D}\left[(fC_{0223}-kC_{0202})v^2-(fC_{2323}+lC_{0202})v\Omega \right. 
\left.+(lC_{0223}+kC_{2323})\Omega^2\right]. \label{48b}
\end{eqnarray}
If the electric and magnetic parts of the Weyl tensor vanish then the Weyl tensor also vanishes and in that case the regularity conditions along the axis are not satisfied as shown in \cite{Celerier}. 

\subsection{Stationary rotating dust case}

 We first investigate the existence of spacetimes which are purely Weyl magnetic (in the sense that their electric part of the Weyl tensor is zero). In that case from \eqref{45b} we have $C_{0303}=0$ and by further substitutions in \eqref{45b} and \eqref{a25} we find first $C_{0101}v^2-2C_{0113}v\Omega+C_{1313}\Omega^2=0$ and then
$\gamma''=2\gamma'/r$ which gives $\gamma'=0$ and finally $\mu=0$. So purely Weyl magnetic non-rigidly rotating cylindrical regular dust spacetimes \eqref{3} do not exist , since the energy density is null.

Now we compute \eqref{48b} using \eqref{16} giving
$$
H_{12}=\frac{\gamma '}{4D} \left[ (fk'-kf')v^2+(fl'-lf')v\Omega+(lk'-kl')\Omega^2\right]
$$
which for the Maitra solution can simply be written (using \eqref{5a}, \eqref{17a}, \eqref{13b}, \eqref{12} and \eqref{f-equ}) as
\begin{equation}
H_{12}= \frac{(k')^3}{8r^2}
\end{equation}
and one can check using \eqref{j5} that $H_{12}\ne 0$ for $r\ne 0$ and $\lim_{r\to 0} H_{12}=0$.

Regarding the fluid rotation, it is already known from \cite{Celerier} that rigidly rotating dust (with $\Omega=0$) necessarily has non-zero vorticity. We will show that this is true also for non-rigid rotation. In the case of the Maitra solution this can be seen by a direct computation using \eqref{ww}, \eqref{j9} which gives \eqref{Omega-exp}
$$
\omega_{\alpha\beta}\omega^{\alpha\beta}=\frac{e^{\gamma}}{2c_1^2(1+x^2)}\ne 0.
$$
For the more general (non-regular) cases with $D=c_2r+c_3$ and $f=1$ we assume, by contradiction, that $\omega_{\alpha\beta}\omega^{\alpha\beta}=0$ (with $\Omega\ne 0$), i.e.
$$
\epsilon(-2DD'\gamma^{\prime})^{1/2}+ \frac{D^2\Omega^{\prime}}{(1+D^2\Omega^2)^{1/2}}=0.
$$
Then substituting $\gamma'$ from \eqref{exp1}  and integrating we get
$\Omega=cD$, with $c$ an integration constant, and then
$
\gamma= -\frac{1}{8} \ln{(1+c^2D^4)}.
$
From \eqref{13b} we get 
$
(k')^2=\frac{c^2(D')^2D^4}{1+c^2D^4}
$
so we can obtain $k$ in terms of hypergeometric functions. Finally to get $l$ we use \eqref{7a} i.e. $l=D^2-k^2$. However, this solution is inconsistent with the Einstein field equations so that one must have $\omega_{\alpha\beta}\omega^{\alpha\beta}\ne 0$. A similar reasoning can be used to check that $H_{12}$ is not identically zero.

Therefore any non-rigidly stationary rotating dust spacetimes \eqref{3} must have non-zero vorticity and non-zero magnetic part of the Weyl tensor.
Concerning future studies, this indicates that the condition of an irrotational fluid can put severe restrictions on cylindrically symmetric rotating spacetime solutions.

\section*{Acknowledgments}

The authors thank the valuable discussions with Dr. Filipe C. Mena.
The author (RC) acknowledges the financial support from FAPERJ (E-26/171.754/2000, E-26/171.533/2002 and E-26/170.951/2006).

\section{Appendix}

For completeness, we present the expressions for the non-zero components of the Weyl tensor $C_{\alpha\beta\gamma\delta}$ in the case of a spacetime metric (\ref{3}). These can also be found in \cite{Celerier}.
\begin{equation}
\aligned
C_{0101}&=-\frac{1}{12}\left[f\gamma^{\prime\prime}-3\left(f\frac{D^{\prime}}{D}-
f^{\prime}\right)\gamma^{\prime}+2f\frac{D^{\prime\prime}}{D} \right. 
\left. -3f^{\prime\prime}+
3\frac{D^{\prime}}{D}f^{\prime}-
2\frac{f}{D^2}(f^{\prime}l^{\prime}+k^{\prime 2})\right], \label{16} \\
C_{0202}&=-\frac{1}{12}\left[f\gamma^{\prime\prime}+3\left(f\frac{D^{\prime}}{D}-f^{\prime}\right)\gamma^{\prime}
-4f\frac{D^{\prime\prime}}{D} \right. \left. +3f^{\prime\prime}-3\frac{D^{\prime}}{D}f^{\prime}+
4\frac{f}{D^2}(f^{\prime}l^{\prime}+k^{\prime 2})\right],  \\
C_{0303}&=-D^2e^{-2\gamma}C_{1212}=
\frac{D^2e^{-\gamma}}{6}\left[\gamma^{\prime\prime}-\frac{D^{\prime\prime}}{D}
+\frac{1}{D^2}(f^{\prime}l^{\prime}+k^{\prime 2})\right],  \\
C_{0113}&=-\frac{1}{12}\left[k\gamma^{\prime\prime}-3\left(k\frac{D^{\prime}}{D}-k^{\prime}\right)\gamma^{\prime}+2k\frac{D^{\prime\prime}}{D} \right. 
\left. -3k^{\prime\prime}+
3\frac{D^{\prime}}{D}k^{\prime}-2\frac{k}{D^2}(f^{\prime}l^{\prime}+k^{\prime 2})\right],  \\
C_{0223}&=-\frac{1}{12}\left[k\gamma^{\prime\prime}+
3\left(k\frac{D^{\prime}}{D}-k^{\prime}\right)\gamma^{\prime}
-4k\frac{D^{\prime\prime}}{D} \right. 
\left. +3k^{\prime\prime}-
3\frac{D^{\prime}}{D}k^{\prime}+4\frac{k}{D^2}(f^{\prime}l^{\prime}
+k^{\prime 2})\right],\\
C_{1313}&=\frac{1}{12}\left[l\gamma^{\prime\prime}-3\left(l\frac{D^{\prime}}{D}-
l^{\prime}\right)\gamma^{\prime}+2l\frac{D^{\prime\prime}}{D}\right.
\left. -3l^{\prime\prime}+3\frac{D^{\prime}}{D}l^{\prime}-
2\frac{l}{D^2}(f^{\prime}l^{\prime}+k^{\prime 2})\right], \\
C_{2323}&=\frac{1}{12}\left[l\gamma^{\prime\prime}+3\left(l\frac{D^{\prime}}{D}-
l^{\prime}\right)\gamma^{\prime}-4l\frac{D^{\prime\prime}}{D} \right.
\left. +3l^{\prime\prime}-3\frac{D^{\prime}}{D}l^{\prime}+4\frac{l}{D^2}(f^{\prime}l^{\prime}+k^{\prime 2})
\right],
\endaligned
\end{equation}
and they satisfy the following relations,
\begin{equation}
\aligned
C_{0101}+C_{0202}&=-\frac{fe^{\gamma}}{D^2}C_{0303}, \\
C_{0113}+C_{0223}&=-\frac{ke^\gamma}{D^2}C_{0303}, \\
C_{1313}+C_{2323}&=\frac{le^\gamma}{D^2}C_{0303}, \\
lC_{0101}+fC_{2323}&=k(C_{0223}-C_{0113}), \label{a25}
\endaligned
\end{equation}
hence we have only three independent components of the Weyl tensor.


\end{document}